\documentclass[showpacs,preprintnumbers,superscriptaddress,amsmath,amssymb]{revtex4-2}
\usepackage[colorlinks=true,bookmarks=false,citecolor=blue,urlcolor=blue]{hyperref}
\usepackage{amsmath,amssymb}
\usepackage{graphicx}
\usepackage{verbatim}
\usepackage{enumitem}
\usepackage{color}
\usepackage[english]{babel}
\usepackage{bm}
\usepackage{natbib}
\usepackage[symbol]{footmisc}
\usepackage{textcomp}

\usepackage{epstopdf}
\usepackage{setspace}
\begin{document}
\title{Degenerate bound states in the continuum in square
and triangular open acoustic resonators}
\author{Almas Sadreev}
\thanks{Corresponding Author}
\email{almas@tnp.krasn.ru}
\author{Evgeny Bulgakov}
\author{Artem Pilipchuk}
\affiliation{Kirensky Institute of Physics Federal Research Center
KSC SB RAS 660036 Krasnoyarsk Russia}
\author{Andrey Miroshnichenko}
\author{Lujun Huang}
\affiliation{School of Engineering and Information Technology,
University of New South Wales Canberra, Northcott Drive, AC,
2600,  Australia}

\begin{abstract}
We consider square and equilateral triangular open acoustic
resonators with the $C_{4v}$ and $C_{3v}$ symmetries,
respectively. There is an unique property of square and triangular
resonators of accidental number four-fold degeneracy of
eigenstates that gives rise to two-fold degenerate
Friedrich-Wintgen (FW) BICs. Compared to usual FW BICs the
degenerate FW BICs maintain high $Q$-factor in wide range of the
size of resonators. That removes the fabrication difficulties of
proper choice of resonator. The presence of degenerate BICs in
triangular resonators is extremely sensitive to switch output
flows by small perturbations with $100\%$ efficiency.

\end{abstract}
\maketitle
\section{Introduction}
A new paradigm for trapping and confining of the resonant modes
have emerged in recent years based on the bound states in the
continuum (BICs) in wave systems. BICs, also known as trapping
mode with infinite large Q-factor, have triggered extensive
interest in photonic and acoustic communities
\cite{Hsu16,Quaranta2018,Huang2020a,Sadreev2021,Joseph2021}. The
most straightforward mechanism of BICs is the symmetrical
incompatibility of closed system states with propagating states of
the continuum
\cite{Bolsterli,Robnik,Schult,Evans1998,Moiseyev2009}. More
interesting, Friedrich-Wintgen (FW) BICs are result of full
destructive interference of two or more resonant modes competing
for leakage into open channels of waveguides
\cite{Friedrich1985,SBR,Sadreev2021}.  The FW BICs can be realized
in open resonators by gradual change of aspect ratio of the
resonator when a degeneracy of eigenfrequencies occur \cite{SBR}.
In acoustic systems, the FW BICs were considered  by many scholars
\cite{Linton2002,Duan2007,Hein2008,Lyapina2015,Huang2020}. The
experimental evidence for the FW BICs was reported by Lepetit and
Kant\'{e} \cite{Lepetit2014} and by Huang {\it et al}
\cite{Huang2021} in the most straightforward configuration of
rectangular resonator opened to attached waveguide.

\begin{figure}[ht!]
\includegraphics[width=0.25\linewidth]{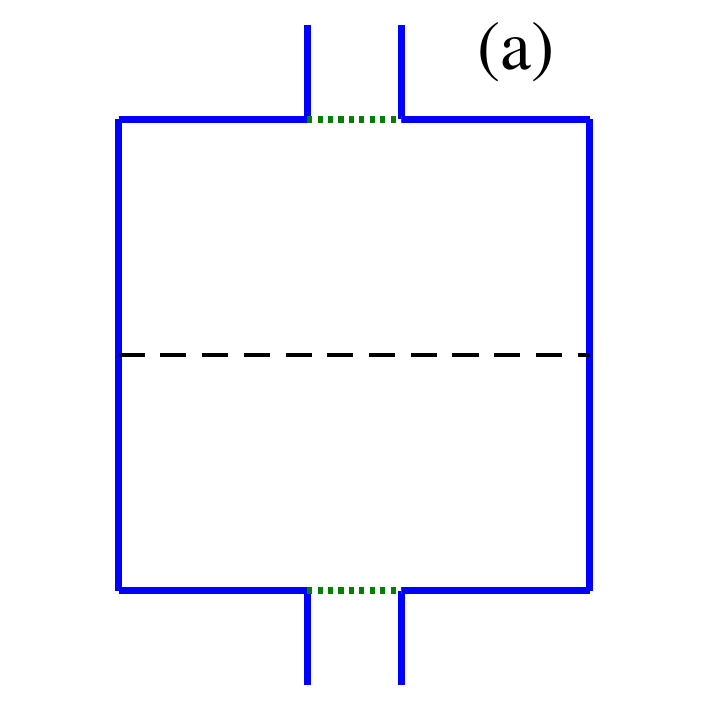}
\includegraphics[width=0.25\linewidth]{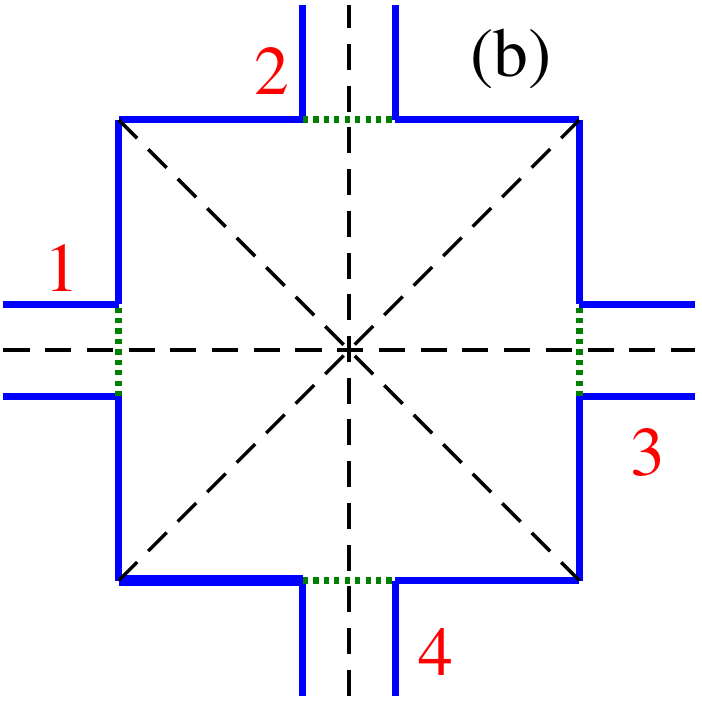}
\includegraphics[width=0.25\linewidth]{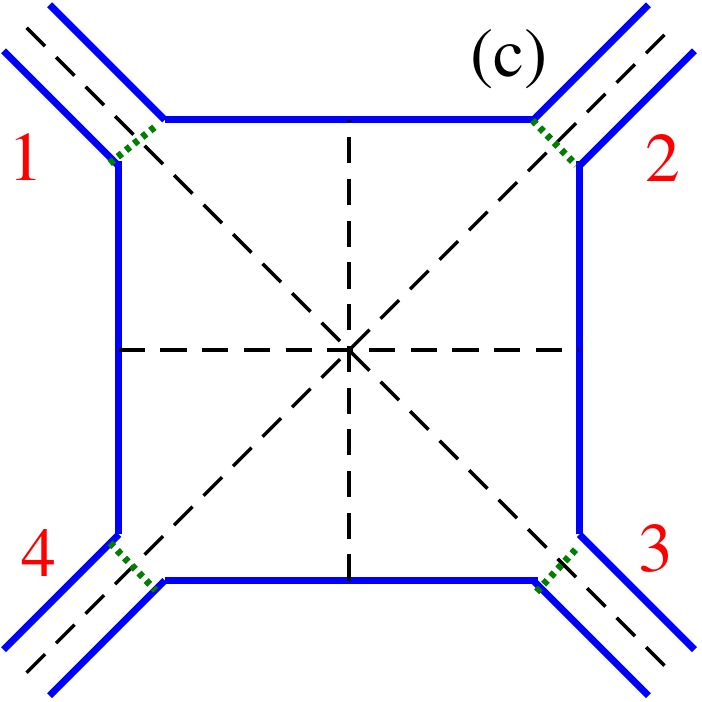}
\includegraphics[width=0.45\linewidth]{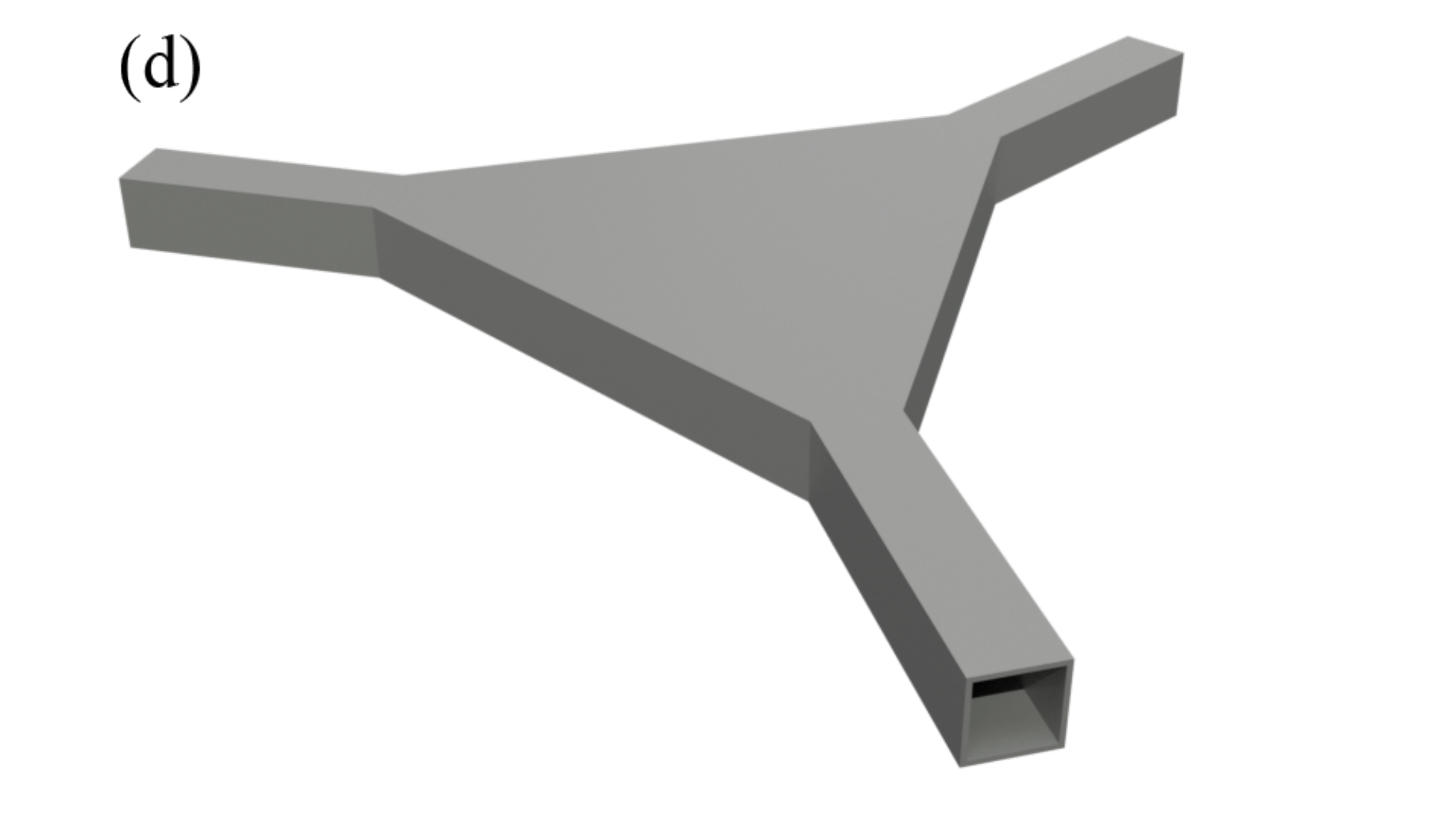}
\caption{(a) Two waveguides attached  to square resonator with the
symmetry $C_{4v}$ lower the symmetry of open system till inversion
symmetry $\sigma$, (b) and (c) Four identical waveguides attached
to square resonator preserve the symmetry $C_{4v}$.  (d) three
waveguides attached to equilateral triangle resonator preserve the
symmetry $C_{3v}$.}
 \label{fig1}
\end{figure}

Although the last time BICs have been successfully demonstrated in
different acoustic resonators fabricated by 3D printing, the
$Q$-factor of quasi-BICs is sensitive to the structure
imperfections. Moreover continual variation of aspect ratio of the
resonator for achievement FW BICs  is very challenging for any
type of resonator, acoustic, metallic or dielectric. Therefore, a
search of FW BICs not sensitive to the aspect ratio of resonators
is highly desired. In the present paper, we advocate square or
equilateral triangular resonators with the group symmetries
$C_{4v}$ and $C_{3v}$, respectively. Its eigenmodes $\psi_{mn}$
and eigenfrequencies $\nu_{mn}^2$ exhibit trivial degeneracy  by
permutation of indices $m\leftrightarrow n$. Here the
eigenfrequencies are given in terms of the frequency $\omega_0=\pi
s/a$, where $s$ is the velocity of sound or light in air, and $a$
is the side size of the resonator. Therefore, one can expect that
opening of the resonator by attachment of waveguides transforms
these degenerate eigenmodes into superradiant mode and FW BIC for
granting as it follows from the FW mechanism of BICs which occur
at degeneracy of eigenfrequencies
\cite{Friedrich1985,Volya,SBR,Sadreev2021}. Variants of attachment
of waveguides to a square resonator are sketched in Fig.
\ref{fig1}. However, for the case (a) in Fig. \ref{fig1} the FW
BICs  occur only after deformation of square resonator into the
rectangular one \cite{Huang2021}. The reason for cancellation of
FW BICs is lowering of  the symmetry $C_{4v}$ of closed square
resonator towards $C_{2v}$ after the (a) case opening.
Quantitatively, the coupling of eigenmodes of a square resonator
with evanescent modes of waveguides perturbs the eigenfrequencies
of the resonator and removes the permutation degeneracy of open
square resonator \cite{Lyapina2018}. Next, consider the cases in
which for opening of the resonators the symmetry $C_{4v}$ is
preserved as sketched in Fig. 1 (b) and (c). In this case the
degenerate eigenmodes $(m,n)$ and $(n,m)$ of closed resonator and
their resonant counterparts of open resonator both are classified
according to the different irreducible representations of the
group $C_{4v}$ and therefore can not be coupled via the continuum
of waveguides that cancels the FW mechanism of the BICs too. The
same consideration is applied to equilateral triangular resonator
with the symmetry group $C_{3v}$.

However, there is a unique case of accidental or number degeneracy
of eigenmodes over two-fold in the square and equilateral triangle
$\nu_{mn}^2=\nu_{m'n'}^2$. For example, in square acoustic
resonator two choices of integers $m=1, n=6$ and $m'=5, n'=4$ have
the same frequency $25/a^2$ \cite{Shaw1974,Kuttler1984}. That
brings four-fold degeneracy of a closed resonator that plays the
key role in the existence of two-fold degeneracy of the FW BICs
provided that open resonators preserve the group symmetries
$C_{4v}$ or $C_{3v}$ as shown in Fig. \ref{fig1} (b)-(d). We show
that the degenerate FW BICs are classified according to the
two-dimensional irreducible representation $E$ of these group
symmetries. Without loss of generality we focus on the acoustic
resonators with Neumann boundary conditions in which BICs can be
directly probed by microphone measurements of acoustic pressure
inside resonators \cite{Huang2020,Huang2021}. Actually, resonators
are three-dimensional, but the acoustic resonators allow
completely disregard the third dimension if the thickness of the
resonator is small enough compared to longitudinal sizes as
sketched in Fig. \ref{fig1} (d). In that case, the solutions are
constant over the third dimension to be excluded from the
solutions presented in Appendix. 
That opens many ways for experimental verifications of theoretical
predictions outlined below in the square and equilateral triangle
resonators.
\section{The SP BICs due to permutation symmetry in open square resonator}
The symmetry of the open resonators is important for
classification and establishment of BICs.  For dielectric
resonators embedded into radiation space  the total symmetry is
given by the structure that determines multipole classification of
radiation and the symmetry of ultra-high $Q$ resonances
(quasi-BICs) of the structure consisted of one or a few
symmetrical dielectric cavities
\cite{Gladyshev2020,Huang2021a,Pichugin2021}. The symmetry of BICs
in  photonic crystals is given by the symmetry of the crystals
\cite{Overvig2020,Overvig2021,Dmitriev2021,Tsimokha2021}. As
different from the above, the symmetry of resonators with attached
directional waveguides is  determined by compatibility of symmetry
of closed resonator with the symmetry of waveguides. The symmetry
of the total system can be lowered or coincide with the symmetry
of closed resonator as illustrated in Fig. \ref{fig1}.

The eigenmodes and eigenfrequencies of the acoustic square
resonator are collected in Appendix. There are two distinct pairs
of degenerate states $\psi_{m,n}, \psi_{n,m}, m \neq n$. The first
pair of the eigenmodes with $m-n$ odd is classified according to
the two-dimensional irreducible representation $E$. The particular
case of $m=2, n=3$ is demonstrated in Fig. \ref{psi23} of
Appendix. The second pair with $m-n$ even is classified according
to the reducible representations. Only the linear combinations
$\psi_{m,n}\pm \psi_{n,m}$ are classified according to the
irreducible one-dimensional representations $B_1$ and $A_1$ or
$B_2$ and $A_2$ respectively. A particular case of eigenmodes
$\psi_{24}$ and $\psi_{42}$ classified according to $B_2$ and
$A_2$ is shown in Fig. \ref{psi24} in Appendix.

Opening of the resonator transforms the real eigenfrequencies of
the closed resonator into the complex eigenfrequencies. The
procedure of transition from closed system onto open system  can
be performed by use of the Feshbach projection technique that
results in the non Hermitian effective Hamiltonian
\cite{Feshbach,Feshbach1962,Rotter1991,Dittes2000,SR2003,Maksimov2015}.
The complex eigenvalues of this Hamiltonian respond for the
position of resonances and resonant line widths and therefore
provide excellent way to establish BICs as eigenmodes of effective
Hamiltonian with real eigenvalues \cite{Sadreev2021}. The specific
form of the effective non Hermitian Hamiltonian is given in
Appendix applied to open acoustic resonators \cite{Maksimov2015}.

Let us first consider the degenerated pair of eigenmodes of closed
square resonator $\psi_{m,n}, \psi_{n,m}$ with $m-n$ odd and
$m\neq n$ classified according to the two-dimensional
representation $E$, for example, the pair presented in Appendix
$\psi_{23}$ and $\psi_{32}$. The coupling matrix elements of these
modes with the first open channel $p=1$ and the second closed
channel $p=2$ for the case in Fig. \ref{fig1} (b) are collected in
Appendix. As a result, we can write the effective non-Hermitian
Hamiltonian projected into these modes whose general form is given
in Eq. (\ref{AHeff}) of Appendix
\begin{equation}\label{Heff23}
    \widehat{H}_{eff}=\left(\begin{array}{cc}
  5/a^2-2ik_1|k_2|\alpha^2+2|k_2|\beta^2 & 0 \cr
  0 & 5/a^2-2ik_1|k_2|\alpha^2+2|k_2|\beta^2 \end{array}\right).
\end{equation}
Here according to Table 2 of Appendix we introduced the notations
$\alpha=W_{2,3;p=1,C=1}=-\frac{2}{\pi}\sin\frac{\pi}{a},
\beta=W_{2,3;p=2,C=1}=f(a)$. The case of Fig. \ref{fig1} (c) gives
the similar diagonal matrix
\begin{equation}\label{Heff23diag}
    \widehat{H}_{eff}=\left(\begin{array}{cc}
  5/a^2-4ik_1|k_2|\gamma^2+4|k_2|\delta^2 & 0 \cr
  0 & 5/a^2-4ik_1|k_2|\gamma^2+4|k_2|\delta^2
\end{array}\right).
\end{equation}
Here the coupling strengths $\gamma$ and $\delta$ can be evaluated
only numerically by integration over thin dotted lines shown in
Fig. \ref{fig1} (c). Irrespectively we obtain that the degenerate
pair of eigenmodes $\psi_{2m,2n+1}$ and $\psi_{2n+1,2m}$
transforms into two degenerate resonances but not BICs. That has
clear physical origin. Since the symmetry of the open resonator
$C_{4v}$ is preserved the eigenmodes of the closed resonator
$\psi_{2,3}$ and $\psi_{3,2}$ are modified but can not be coupled
through the open continuum $p=1$ of waveguides owing to the
symmetry $C_{4v}$ of the open square resonator. As a result the FW
mechanism of BICs is cancelling.

Next, we consider the pair $\psi_{s,a}=\psi_{m,n}\pm \psi_{n,m},
m\neq n$ and $m-n$ even which are classified according to the
one-dimensional irreducible representations $B_2$ and $A_2$.
Example of these eigenfunctions $m=2, n=4$ are illustrated in Fig.
\ref{psi24} of SI. For the case in Fig. \ref{fig1} (b) all
coupling matrix elements with the first open channel of each
waveguide equal zero as clearly seen in \ref{psi24} of SI. As a
result only closed channels of waveguides contribute in the
effective Hamiltonian which in the space of the eigenmodes
$\psi_{2,4}$ and $\psi_{4,2}$ takes the following form
\begin{equation}\label{Heff24}
    \widehat{H}_{eff}=\left(\begin{array}{cc}
  10/a^2+4|k_2|(b(a)^2+c(a)^2) & -4|k_2|b(a)c(a) \cr
  -4|k_2|b(a)c(a) & 10/a^2+4|k_2|(b(a)+c(a)^2)\end{array}\right).
\end{equation}
Therefore we have  two SP BICs $\psi_{s,a}$ shown in Fig. 2 of SI
with the eigenfrequencies
\begin{equation}\label{BIC24d}
   \nu_{s,a}^2=10/a^2+4|k_2|\left\{\begin{array}{l}b(a)^2 \\
   c(a)^2 \end{array}\right.
\end{equation}
where the coupling constants $b(a), c(a)$ are collected in Eq.
(\ref{Wp2}) of SI.

\section{The two-fold degenerate BICs in open square resonator}

As it was said in the Introduction there is the accidental number
degeneracy of two doublets each degenerated by permutation of
indices. For example, eigenmodes with indices $1,6$ and $4,5$,
both classified according to the two-dimensional irreducible
representation $E$ have the same eigenvalue
$\nu_{16}^2=\nu_{4,5}^2=25/a^2$. The next quartet of degenerated
eigenmodes is, for example, $3,10$, $10,3$ and $7,8$, $8,7$ with
the eigenvalue $85/a^2$. It is reasonable to project the effective
non-Hermitian Hamiltonian onto this space of the eigenmodes. Let
us enumerate the eigenmodes as follows
$$\phi_1=\psi_{1,6}, \phi_2=\psi_{5,4}, \phi_3=\psi_{6,1},
\phi_4=\psi_{4,5}.$$ In order to close the second channel $p=2$ of
waveguides, we consider the eigenfrequencies  of the resonator
below the second cutoff of the waveguide, i.e., $\nu_{m,n}/a<1$ if
to express side sizes of resonators $a$ via the width of
waveguides $d=1$. In particular, for the eigenstates under
consideration the size of resonator is to exceed $a>5$.

Let us, first, consider the case of sidewall connection of
waveguides as shown in Fig. \ref{fig1} (b). The coupling matrix
elements of these eigenmodes with open channel $p=1$ and closed
channel $p=2$ are collected in Tables \ref{Tab2} and \ref{Tab3} of
Appendix. As a result, we obtain for the effective Hamiltonian in
truncated Hilbert space of the eigenfunctions $\phi_j, j=1,2,3,4$
\begin{equation}\label{Heffside}
    \widehat{H}_{eff}=\frac{1}{a^2}\left(\begin{array}{cc}
  \hat{h}_{eff} & 0 \\
  0 & \hat{h}_{eff} \\
\end{array}\right), \widehat{h}_{eff}=\left(\begin{array}{cc} \epsilon-i\gamma_1 &
-u-i\sqrt{\gamma_1\gamma_2}\cr -u-i\sqrt{\gamma_1\gamma_2}
&-\epsilon-i\gamma_2\end{array}\right),
\end{equation}
where according to Tables 2 and 3 In Appendix
\begin{equation}\label{heffME}
\epsilon=|k_2|(g(a)^2-h(a)^2), u=2|k_2|g(a)h(a),
\gamma_1=\frac{4k_1}{a^2},
\gamma_2=\frac{2k_1}{\pi^2}\sin^2\left(\frac{2\pi}{a}\right),
\end{equation}
and according to Eq. (4) in Appendix $k_1=\nu=\nu_{16}=5/a,
|k_2|=7/a$.

\begin{figure}[ht]
\includegraphics[width=0.275\linewidth]{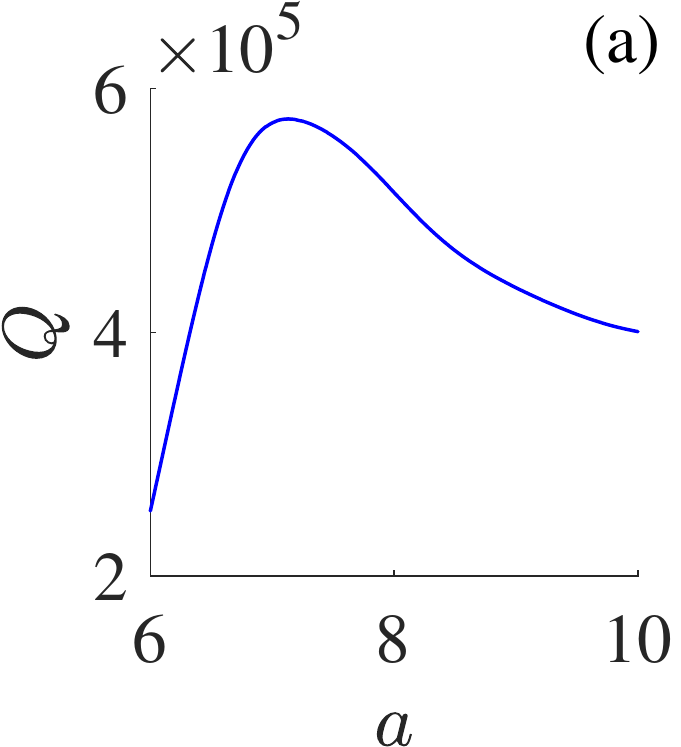}
\includegraphics[width=0.3\linewidth]{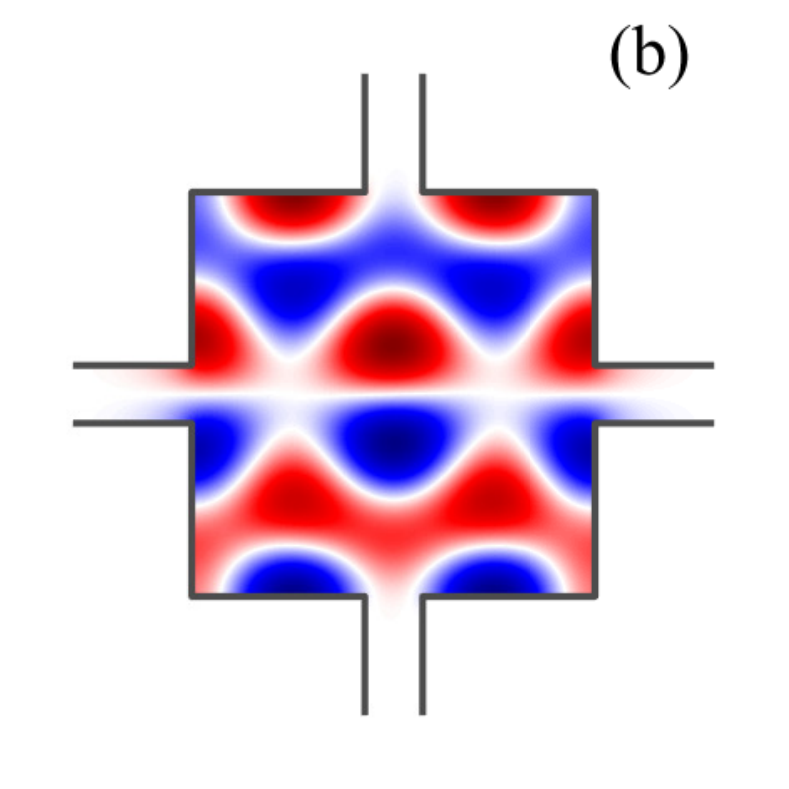}
\includegraphics[width=0.3\linewidth]{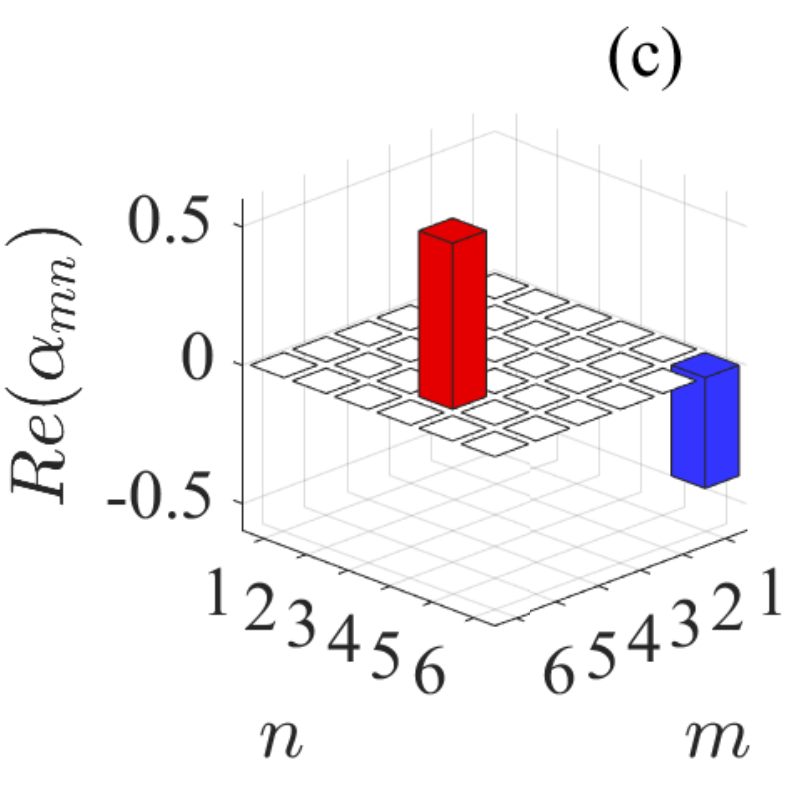}
\caption{(a) The $Q$-factor of quasi FW BIC shown in subplot  (b)
with the modal expansion coefficients shown in (c). } \label{fig2}
\end{figure}

One can see that the effective Hamiltonian consists of two
identical blocks $2\times 2$, each of them has typical form for
description of FW BICs \cite{Friedrich1985,Volya,Sadreev2021}.
Comparison with expressions (\ref{Heff23}) and (\ref{Heff24})
shows that the eigenmodes $(1,6)$ and $(5,4)$ and, respectively,
$(6,1)$ and $(4,5)$ are coupled through the continuum of
waveguides that gives rise to FW mechanism of two degenerate BICs
classified according to the two-dimensional irreducible
representation $E$. If we neglected by closed channel, $p=2$ one
could have two degenerate FW BICs for $a>5$. However, the
evanescent modes of waveguides play the principal role because
they give rise to real coupling $u$ between modes. For $u\neq 0$,
the FW BIC occurs in the framework of the Hamiltonian
(\ref{Heffside}) according to the following equation
\cite{Volya,Kikkawa2019,Sadreev2021}
\begin{equation}\label{u_eps}
u(\gamma_1-\gamma_2)=2\epsilon\sqrt{\gamma_1\gamma_2}.
\end{equation}
It is easy to fulfill  Eq. (\ref{u_eps}) if the parameters $u,
\gamma_1, \gamma_2, \epsilon$ were  independent. However, for the
present case all constants  depend on only the square size $a$ as
given in Eq. (\ref{heffME}). As Comsol MultiPhysics shows in Fig.
\ref{fig2} (a)  the condition (\ref{u_eps}) is not fulfilled for
variation $a$ that defines the solution as quasi FW BIC although
with extremely large $Q$-factor around 600000. In view material
losses of 3d printed acoustic resonators which restrict the $Q$
factor by order of $10^3$ \cite{Huang2021} one can consider the
solution in Fig. \ref{fig2} (b) as the FW BIC. Remarkably,
compared to usual FW BICs with sharp peak in the $Q$-factor for
variation of the sizes of resonator the degenerate FW BICs
maintain extremely high $Q$-factor in wide range of the size. That
relieves experimentalists from the fabrication difficulties of
proper choice of the resonators.

Fig. \ref{fig2} (c) clearly shows that this BIC is composed of
eigenmodes $(1,6)$ and $(5,4)$. The FW BIC composed of the
eigenmodes $(6,1)$ and $(4,5)$ differs from this solution by
$90^o$ rotation.

\begin{figure}[ht]
\includegraphics[width=0.3\linewidth]{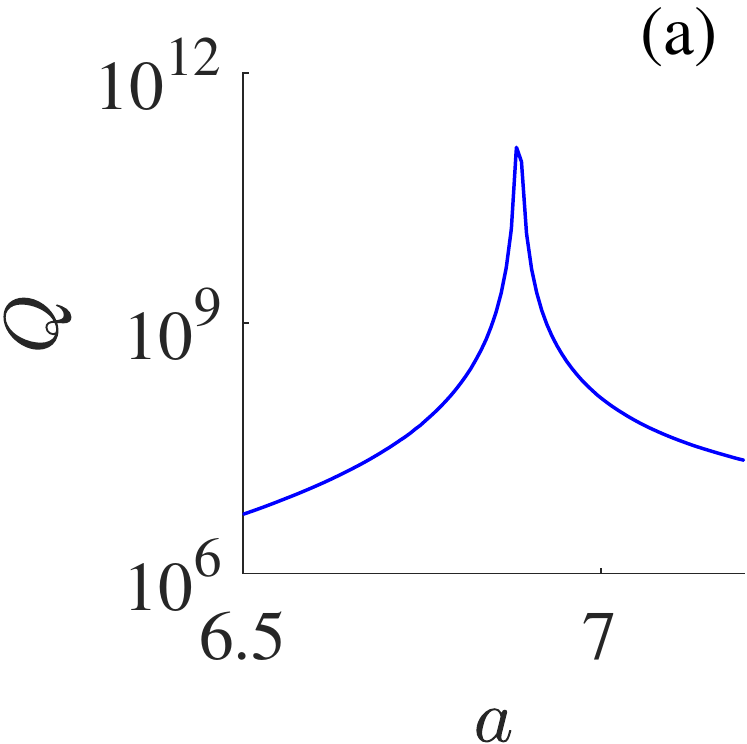}
\includegraphics[width=0.3\linewidth]{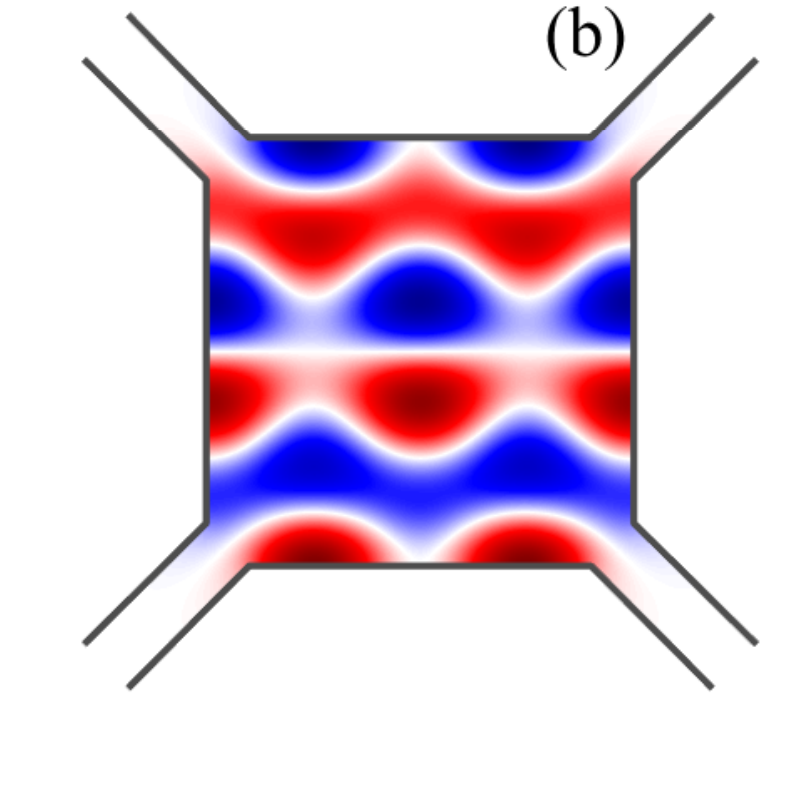}
\includegraphics[width=0.3\linewidth]{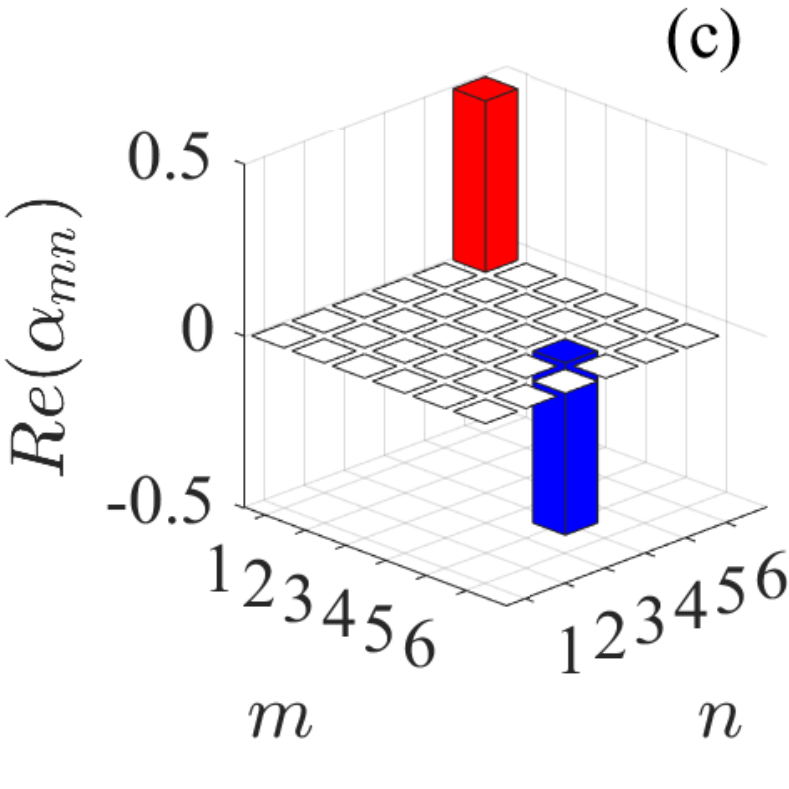}
\caption{The same as in Fig. \ref{fig2} but for vertex connection
of resonator to waveguides as shown in Fig. \ref{fig1} (c).}
\label{fig3}
\end{figure}

In order to restore true degenerate FW BICs, we attach waveguides
at the vertices of the square as shown in Fig. \ref{fig1} (c)
which also give equal couplings of eigenmodes of the square
resonator with the continua of waveguides. However similar to side
coupling of square with waveguides the evanescent modes remove the
number degeneracy of modes $1,6$ and $5,4$. Numerics reveals two
degenerate FW BICs with the frequency $\nu = 0.832051$ occur for
variation of square size $a=6.8818$, the first of which is shown
in Fig. \ref{fig3}. One can see that FW BIC is superposed of two
eigenmodes $(1,6)$ and $(5,4)$.
The second degenerate FW BIC is obtained by $90^o$ rotation and
both BICs are classified according to the two-dimensional
representation $E$.

\section{Degenerate BICs in equilateral triangle}

\begin{figure}[ht]
\includegraphics[width=0.6\linewidth,clip=]{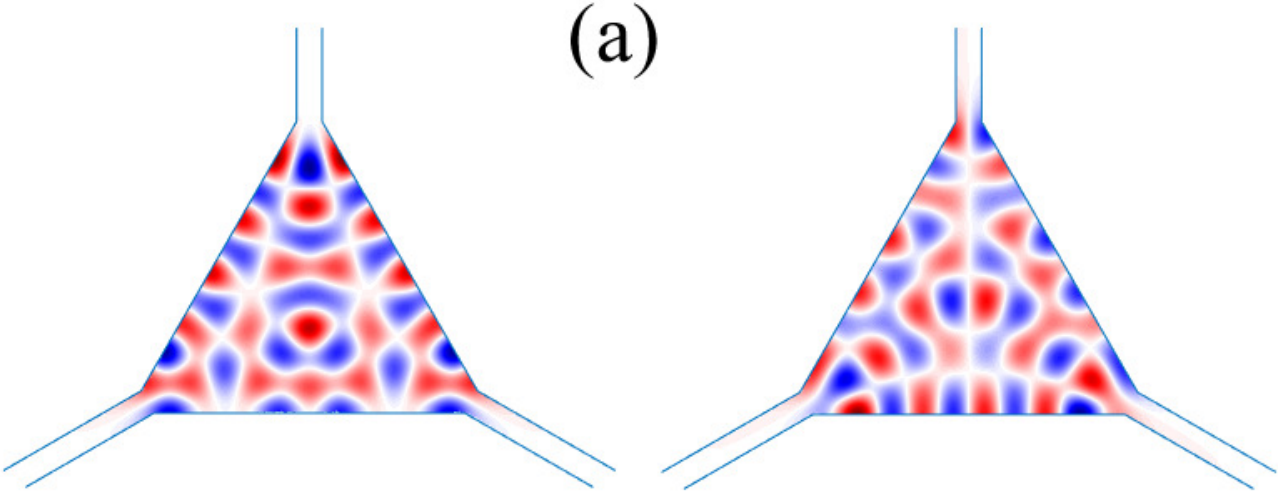}
\includegraphics[width=0.3\linewidth,clip=]{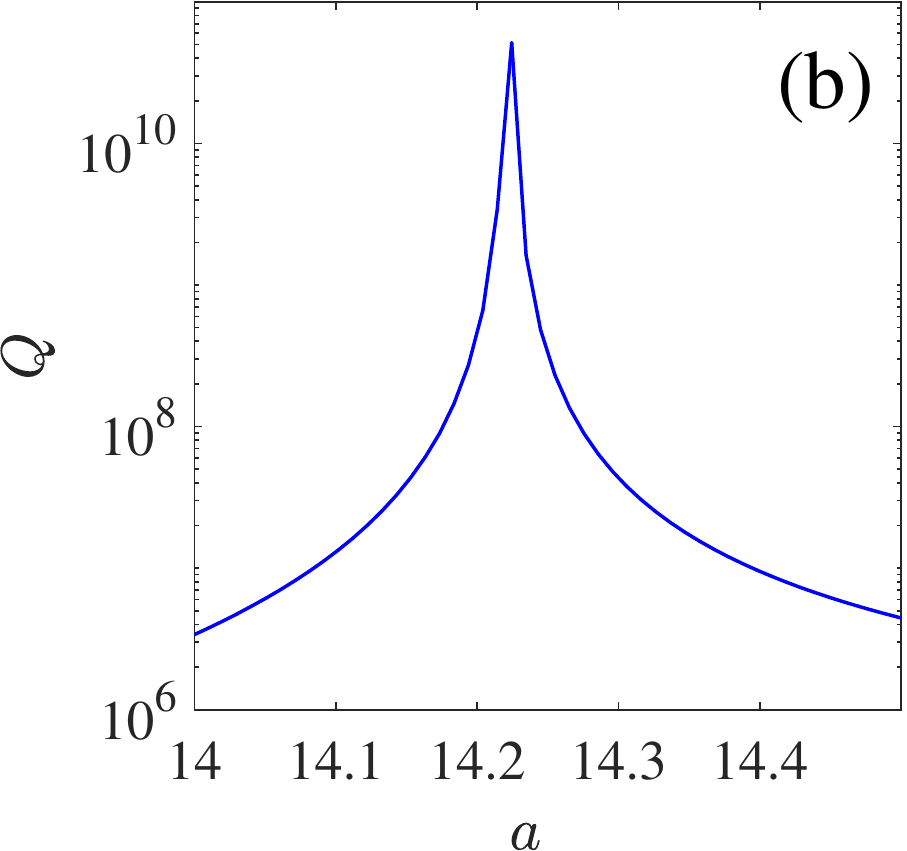}
\caption{(a) Degenerate BICs with the frequency $\nu_{BIC}=0.8938$
classified according to the two-dimensional irreducible
representation $E$ of group symmetry $C_{3v}$. $a=a_c=14.1939$.
(b) The $Q$-factor dependence on side size of triangle $a$. }
\label{fig4}
\end{figure}

Less obvious cases of the number degeneracy exist in the
equilateral triangle with the eigenmodes and eigenfrequencies
presented in Appendix. Similar to a square resonator, all
eigenmodes are two-fold degenerate relative to $m\leftrightarrow
n$, giving rise to the  FW BICs. However there are also
exceptional cases of the four-fold accidental number degeneracy,
for example, for $m=-11, n=-19$ and $m=-16, n=-17$ . Pressure
profiles of these four eigenmodes with the lowest frequency
$\nu\approx 40$ are shown in Appendix. Respectively, with opening
of the triangular resonator with three attached waveguides these
four eigenmodes are transformed into two superradiant modes and
two FW BICs classified according to the two-dimensional
irreducible representation $E$ of group symmetry $C_{3v}$. Similar
to open square resonator, only vertex attachment of waveguides
allows existence of degenerate BICs, as shown in Fig. \ref{fig4}.
The resonant eigenmodes of the open triangular resonator can be
found only numerically as distinctive from the case of the square
open resonator.

\begin{figure}[ht]
\includegraphics[width=0.45\linewidth,clip=]{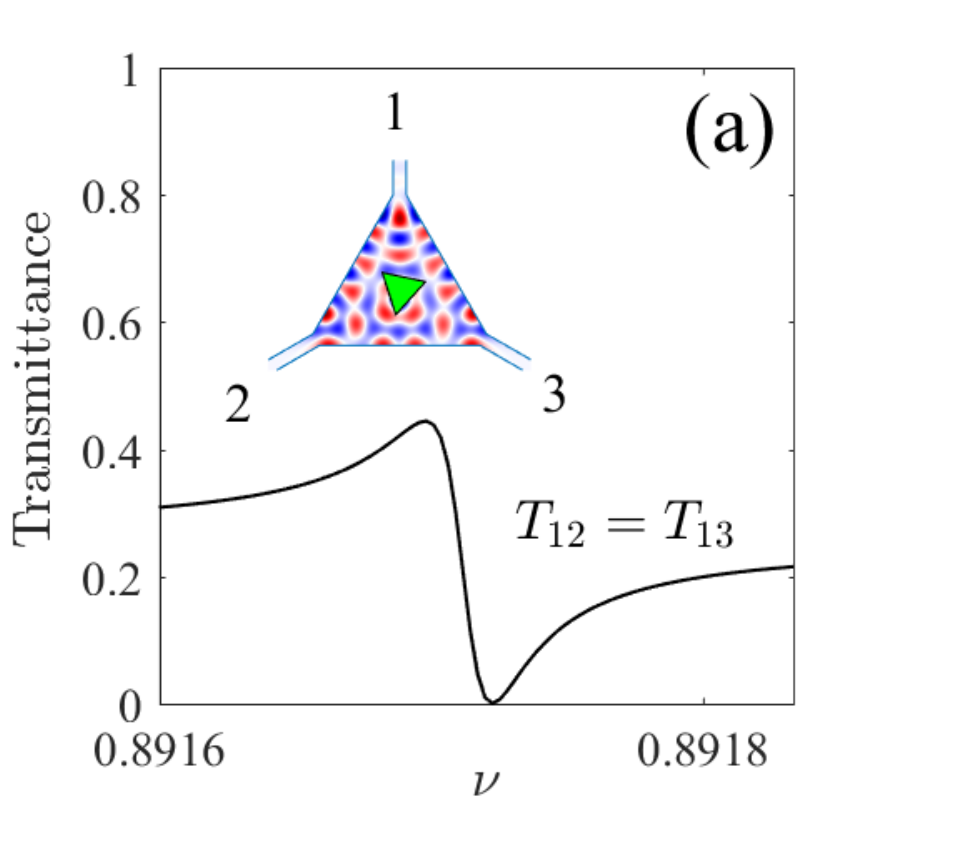}
\includegraphics[width=0.45\linewidth,clip=]{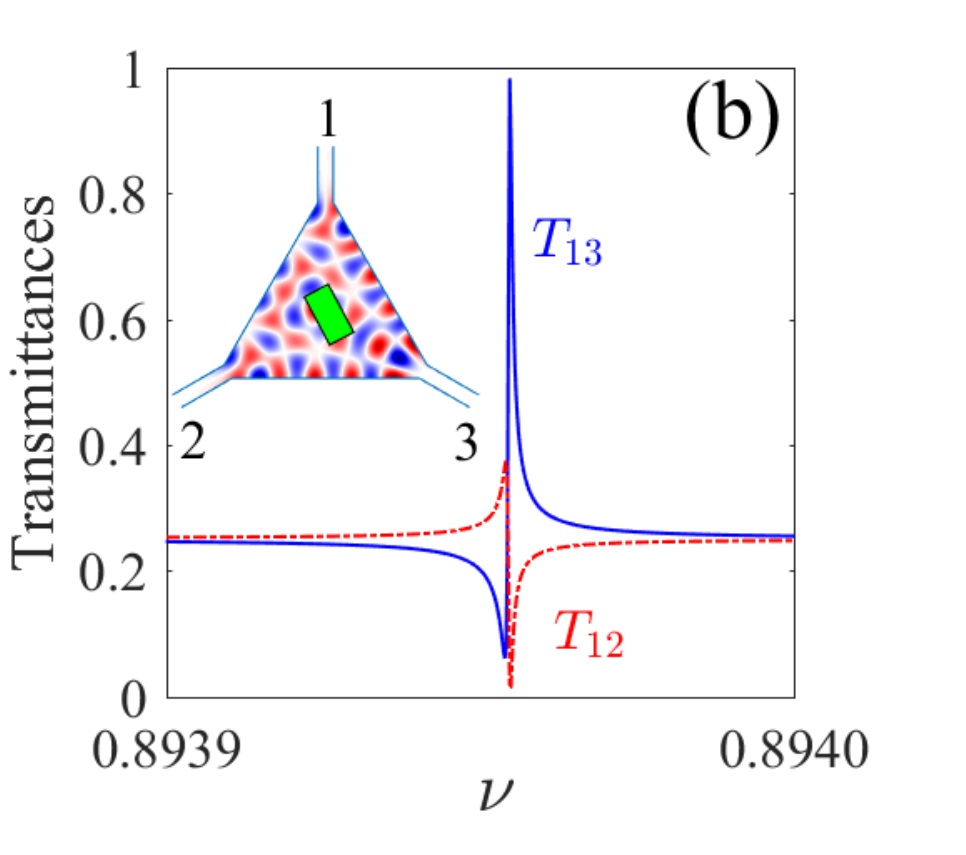}
\caption{Transmittance from input waveguide 1 into the output
waveguides 2 and 3 in the triangular resonator with rotated (a)
triangular pencil  and (b) rectangular pencil. The frequency of
input wave is tuned onto the frequency of degenerate FW BICs shown
in Fig. \ref{fig4}. } \label{fig5}
\end{figure}

The existence of degenerate resonances and, in particular, BICs
opens a way of highly effective manipulating out power flows by
small perturbations. In particular, that can be done by slight
violation of symmetry of the system, say, by slight local pressure
onto the resonator walls. In the present letter, we apply a local
perturbation in the form of pencil of triangular or rectangular
cross-sections at the center of the resonator which can be rotated
by the angle $\theta$ relative to the resonator.  Fig. \ref{fig5}
demonstrates the striking result of the symmetry incompatibility
of the resonator and perturbation. Fig. \ref{fig5} (a) shows that
perturbation which preserves the symmetry $C_{3}$ of the total
system can not manipulate by acoustic flows irrespective of the
angle $\theta$. While the perturbation of the symmetry $C_2$ whose
symmetry is not compatible with the symmetry of the total system
drastically changes the output acoustic flows.

\section{Conclusions}
We considered square and triangular resonators with the symmetries
whose eigenmodes are classified to the irreducible representations
of group symmetries $C_{4v}$ and $C_{3v}$ among of which there is
the two-dimensional representation $E$. One can preserve the
symmetries of open resonators owing to proper connection of
waveguides to the resonator as sketched in Fig. \ref{fig1}. The
Hamiltonian of closed resonator $H_B$ transforms into
non-Hermitian effective Hamiltonian $H_{eff}$ after the Feshbach
projection of the total system into space of eigenfunction of
$H_B$ \cite{Feshbach}. Respectively, the eigenmodes of $H_B$ with
real eigenvalues are substituted by resonant modes, which are the
eigenmodes of $H_{eff}$ with complex eigenvalues. Both
Hamiltonians commute with the symmetry group transformations, and
therefore one can expect that there are two-fold degenerate
resonant states classified according to the two-dimensional
irreducible representation $E$. However, analytical consideration
in Section 3 explicitly shows these resonant states can not be
true FW BICs because of the absence of interaction through the
continuum of waveguides. In order to realize the degenerate FW
BICs, we explore the unique property of square and triangular
resonators of numeric accidental degeneracy of eigenmodes. As a
result, we obtain the four-fold degeneracy of eigenmodes of $H_B$
which transform into the two-fold degenerate FW BICs and two
superradiant resonances. Shaw has presented even more unique cases
of number eight-fold degeneracy for $m=5, n=34; m=10, n=33; m=13,
n=32; m=24, n=25$ with higher eigenfrequencies \cite{Shaw1974}.
Respectively, we can expect the four-fold degenerate FW BICs or go
above the first cutoff $\nu=1$ towards the FW BICs embedded into a
few continua of the next propagating bands of waveguides
\cite{Remacle1990}. In general, degenerate BICs can also occur in
open systems symmetrical, for example, relative to axial
rotations. Then the Hilbert space of total system splits into a
direct sum of spaces specified by the azimuthal index $m$.
Respectively, the BICs, if they exist, are degenerate relative to
$\pm m$  because of time-reversal symmetry. Examples of such
degenerate BICs were reported in the periodical array of
dielectric spheres \cite{Bulgakov2015} and disks
\cite{Bulgakov2017}. However, these BICs are not degenerate in
each Hilbert subspace specified by the azimuthal index $m$.

\section{Appendix}
\subsection{The eigenmodes of square resonator classified according
to irreducible representations of square symmetry group $C_{4v}$}

In the acoustic square resonator with the Neumann boundary
conditions result in the following eigenmodes
\begin{equation}\label{eigfreqrect}
    \psi_{m,n}(x,y) = \sqrt{\frac{(2-\delta_{m,1})(2-\delta_{n,1})}{a} }
    \cos\left(\frac{\pi (m-1)x}{a}\right)\cos\left(\frac{\pi (n-1)y}{a}\right)
\end{equation}
with the eigenfrequencies
\begin{equation}\label{eigfreqrect}
    \nu_{m,n}^2=\omega_{m,n}^2/\omega_0^2=(m-1)^2+(n-1)^2,
    m,n=1,2,3,\ldots,
    \end{equation}
    where $\omega_0=\pi s/a$, $s$ is the velocity of sound in air and
    $a$ is the size of square.

\begin{figure}[ht]
\includegraphics[width=12cm,clip=]{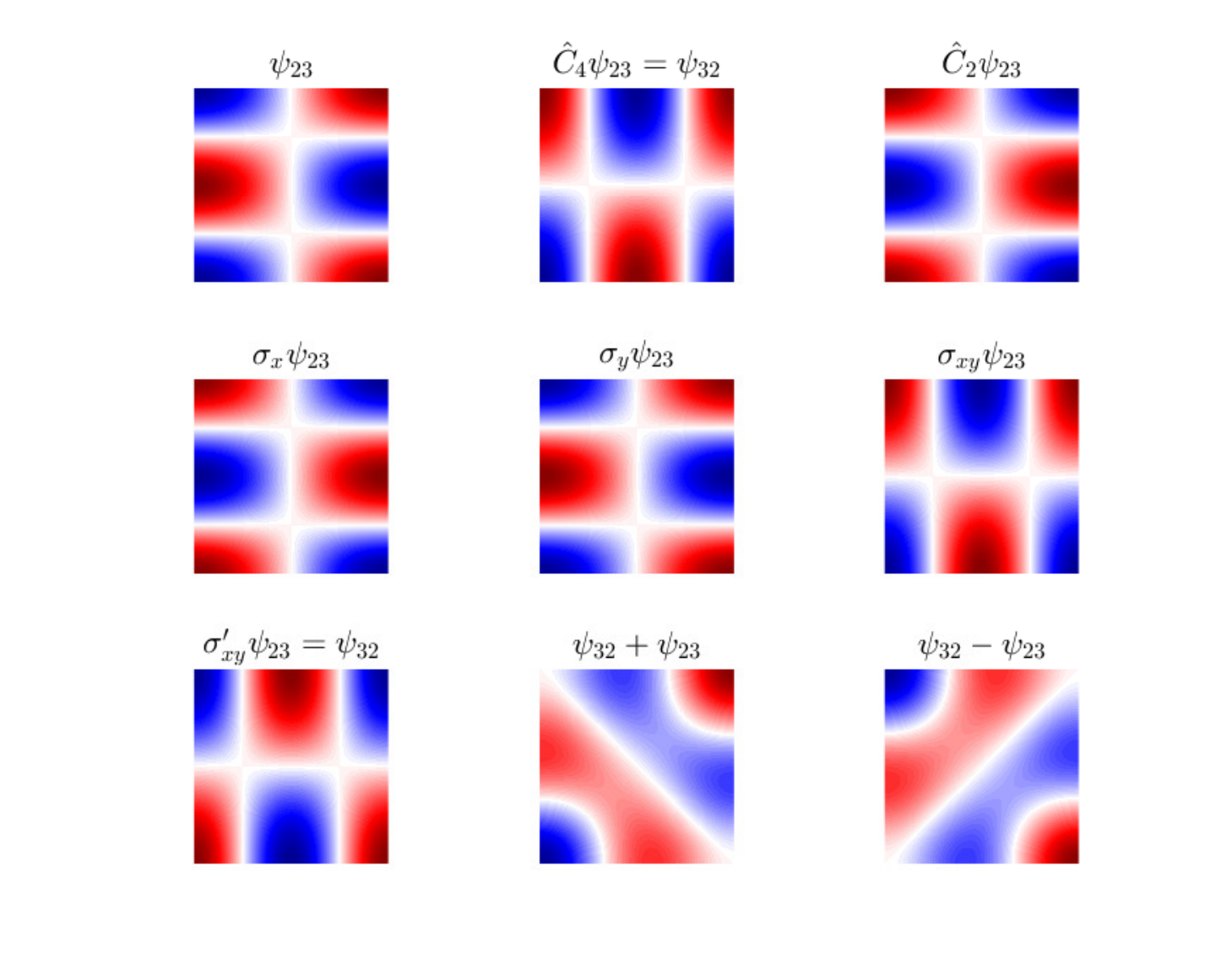}
\caption{Symmetry group $C_{4v}$ transformations of the eigenmode
$\psi_{23}(x,y)$.} \label{psi23}
\end{figure}

\begin{figure}[ht]
\includegraphics[width=12cm,clip=]{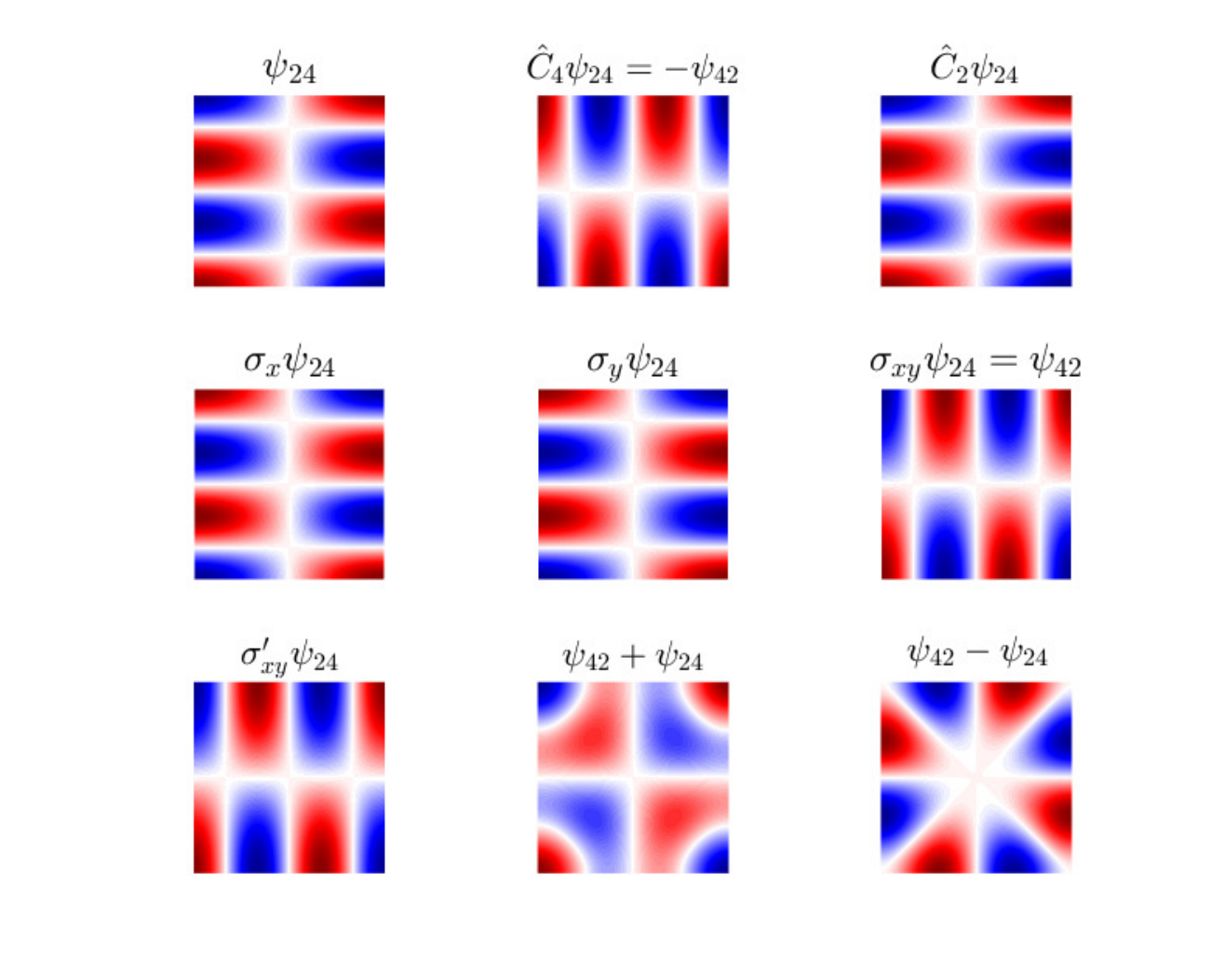}
\caption{Symmetry group $C_{4v}$ transformations for the eigenmode
$\psi_{24}(x,y)$.} \label{psi24}
\end{figure}

\begin{figure}[ht]
\includegraphics[width=12cm,clip=]{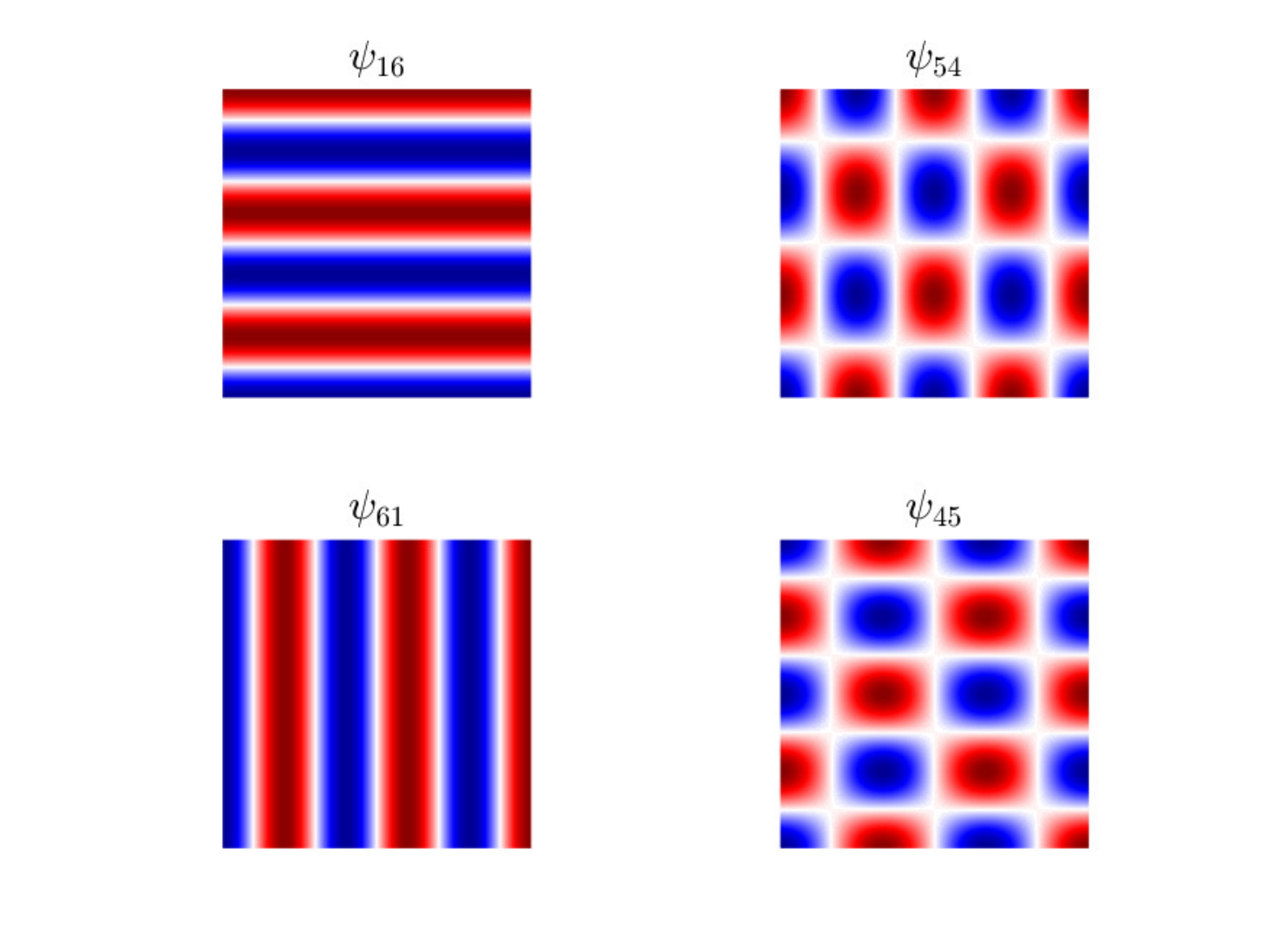}
\caption{The number degenerated eigenfunctions of square.}
\label{psi16_54}
\end{figure}

\begin{table}
\caption{The characters of irreducible representations of group
symmetry $C_{4v}$}
\begin{tabular}{|c|c|c|c|c|c|c|}
  \hline
$C_{4v}$ & 1 & $C_2$ & $2 C_4$ &$\sigma_x, \sigma_y$ &
$\sigma_{xy}, \sigma'_{xy}$ & basic modes \\  \hline
$A_1$   &1  & 1     & 1       &1 &  1                  & $z$ \\
  \hline
$A_2$   & 1 & 1     & 1       & -1& -1 &$J_z$ \\
  \hline
$B_1$ & 1 & 1 & -1 & 1& -1& $x^2-y^2$\\
  \hline
$B_2$ & 1 & 1 & -1 &-1& 1& $xy$ \\
  \hline
$E$ & 2 & -2 & 0 & 0& 0 &$x, y$\\
  \hline
 \end{tabular}
 \label{Tab1}
\end{table}
 The group of symmetry $C_{4v}$ of square consists of rotations $C_4$ and $C_2$, two
mirror reflections $\sigma_v$ along the square axis $x$ and $y$
and $\sigma_v'$ along the diagonals of square
\cite{Landau58,Bir_Pikus}. Tabl. \ref{Tab1} shows the irreducible
representations of each symmetry transformation and their
characters \cite{Bir_Pikus}. Fig. \ref{psi23} shows that the
eigenmodes $\psi_{m,2n+1}$ and $\psi_{2n+1,m}$ belong to the
two-dimensional irreducible representation $E$. However the
eigenmodes $\psi_{2m,2m+2n}$ and $\psi_{2m+2n,2m}$ are classified
by the reducible representations. As Fig. \ref{psi24} shows only
the linear combinations $\psi_{2m+2n,2m}+\psi_{2m,2m+2n}$ and
$\psi_{2m+2n,2m}-\psi_{2m,2m+2n}$ are classified according to the
irreducible one-dimensional representations $B_2$ and $A_2$
respectively. Similarly, the linear combinations
$\psi_{2m+2n+2,2m+1}+\psi_{2m+1,2m+2n+2}$ and
$\psi_{2m+2n+2,2m+1}-\psi_{2m+1,2m+2n+2}$ are classified according
to the irreducible one-dimensional representations $B_1$ and $A_1$
respectively.
In Fig. \ref{psi16_54} we present an example of eigenmodes
$\psi_{16}$ and $\psi_{54}$ degenerated accidentally
\cite{Shaw1974,Kuttler1984}.

\subsection{Effective non Hermitian Hamiltonian} The procedure of
the Feshbach projection of the total Hilbert space of the total
system closed resonator plus waveguides with Neumann boundary
conditions onto Hilbert space of the eigenmodes of closed
resonator is described in Ref. \cite{Maksimov2015}. In application
to the square acoustic resonator we have
\begin{equation}\label{AHeff}
   \widehat{H}_{eff}=\nu^2_{mn}\delta_{mm'}\delta_{nn'}-\sum_{p=1}^{\infty}
   \sum_{C}ik_p\widehat{W}_{Cp}\widehat{W}_{Cp}^{\dagger},
\end{equation}
where $k_p$ are the propagating momenta of the $p$-th channel in
waveguides of width $d$. In what follows all dimensions are
measured in terms of $d$, i.e., $d=1$. Then
\begin{equation}\label{disp}
    \nu^2=k_p^2+(p-1)^2, p=1,2,3,\dots
\end{equation}
with
\begin{equation}\label{chanel modes}
    \phi_p(x,y)=\sqrt{2-\delta_{p,1}}\cos(\pi(p-1)y)e^{i k_p x}.
\end{equation}

The index $w$ in Eq. (\ref{AHeff}) sorts waveguides. For
attachments of waveguides shown in Fig. 1 of the main text the
coupling matrix $\widehat{W}_{Cp}$ can be evaluated analytically
\begin{eqnarray}\label{W}
    &W_{mn;p,C=1,3}=\int_{-1/2}^{1/2}\psi_{mn}(x=\mp
    a/2,y)\phi_p(y)dy,&\\
&W_{mn;p,C=2,4}=\int_{-1/2}^{1/2}\psi_{mn}(x,y=\pm
    a/2)\phi_p(x)dx.&
\end{eqnarray}
After integration in geometry shown in Fig. \ref{fig1} (b) of the
paper we obtain for the waveguide 1:
\begin{eqnarray}\label{WmnL}
    &W_{mn;p,C=1}=\frac{\sqrt{(2-\delta_{m,1})(2-\delta_{n,1})(2-\delta_{p,1})}}
    {\pi}
[\frac{\sin[\frac{\pi}{2a}(n-1+a(p-1))]
\cos[\frac{\pi}{2}(n-1+a(p-1))]}{n-1+a(p-1)}&\nonumber\\
&+\frac{\sin[\frac{\pi}{2a}(n-1-a(p-1))]
\cos[\frac{\pi}{2}(n-1-a(p-1))]}{n-1-a(p-1)}\}.&
\end{eqnarray}
Some particular coupling matrix elements relevant for the paper
are collected in Tables \ref{Tab2} and \ref{Tab3}
\begin{table}
        \centering
                \caption{Coupling matrix elements with the first open channel $p=1$}
                \label{Tab2}
\begin{tabular}{c|c|c|c|c} \hline\hline
modes $m,n$ $\setminus C$ &1 &2 &3 &4 \\
\hline
$2,3$ & -$\frac{2}{\pi}\sin\frac{\pi}{a}$ & 0 & $\frac{2}{\pi}\sin\frac{\pi}{a}$ & 0  \\
$3,2$ & 0 &$\frac{2}{\pi}\sin\frac{\pi}{a}$ & 0 & $-\frac{2}{\pi}\sin\frac{\pi}{a}$   \\
\hline
$2,4$ & 0 & 0 & 0 & 0  \\
$4,2$ & 0 & 0 & 0 & 0   \\
\hline
$1,6$ & 0 &$\frac{\sqrt{2}}{a}$& 0 &   -$\frac{\sqrt{2}}{a}$ \\
$5,4$ & 0 & $\frac{1}{\pi}\sin\frac{2\pi}{a}$  & 0 & -$\frac{1}{\pi}\sin\frac{2\pi}{a}$\\
$6,1$ & $\frac{\sqrt{2}}{a}$ & 0 & -$\frac{\sqrt{2}}{a}$& 0\\
$4,5$ & $\frac{1}{\pi}\sin\frac{2\pi}{a}$ & 0 & -$\frac{1}{\pi}\sin\frac{2\pi}{a}$ & 0\\
\hline \hline
\end{tabular}
\end{table}
\begin{table}
        \centering
                \caption{Coupling matrix elements with the second closed channel $p=2$}
                \label{Tab3}
\begin{tabular}{c|c|c|c|c} \hline\hline
modes $m,n$ $\setminus C$ &1 &2 &3 &4 \\
\hline
$2,3$ & 0 &-$f(a)$ & 0 & $f(a)$  \\
$3,2$ & -$f(a)$ & 0 & $f(a)$ & 0   \\
\hline
$2,4$ & -$b(a)$ & -$c(a)$ & $b(a)$ & $c(a)$  \\
$4,2$ & $c(a)$ & $b(a)$ & -$c(a)$ & -$b(a)$   \\
\hline
$1,6$ & $g(a)$ & 0         & $g(a)$    & 0 \\
$5,4$ &-$h(a)$ & 0         & -$h(a)$   & 0 \\
$6,1$ &  0     & $g(a)$    & 0         & $g(a)$\\
$4,5$ &  0     & $h(a)$    & 0         & $h(a)$\\
\hline \hline
\end{tabular}
\end{table}
where according to Eq. (\ref{WmnL}) we denote
\begin{eqnarray}\label{Wp2}
&b(a)=\frac{4\sqrt{2}a}{\pi(a^2-9)}\sin\frac{\pi a}{2}\cos\frac{3\pi}{2a},&\nonumber\\
&c(a)=\frac{4\sqrt{2}a}{\pi(a^2-4)}\sin\frac{\pi a}{2}\cos\frac{\pi}{2a},&\nonumber\\
    &f(a)=\frac{4\sqrt{2}a}{\pi(a^2-4)}\cos\frac{\pi}{a}\cos\frac{\pi a}{2},&\nonumber\\
&g(a)=-\frac{4a}{\pi(a^2-25)}\cos\frac{5\pi}{2a}\sin\frac{\pi a}{2},&\nonumber\\
&h(a)=\frac{4\sqrt{2}a}{\pi(a^2-16)}\cos\frac{2\pi}{a}\cos\frac{\pi
a}{2}.&
    \end{eqnarray}

\subsection{Eigenmodes of equilateral triangular billiard} 
\begin{figure}[ht]
\includegraphics[width=10cm,clip=]{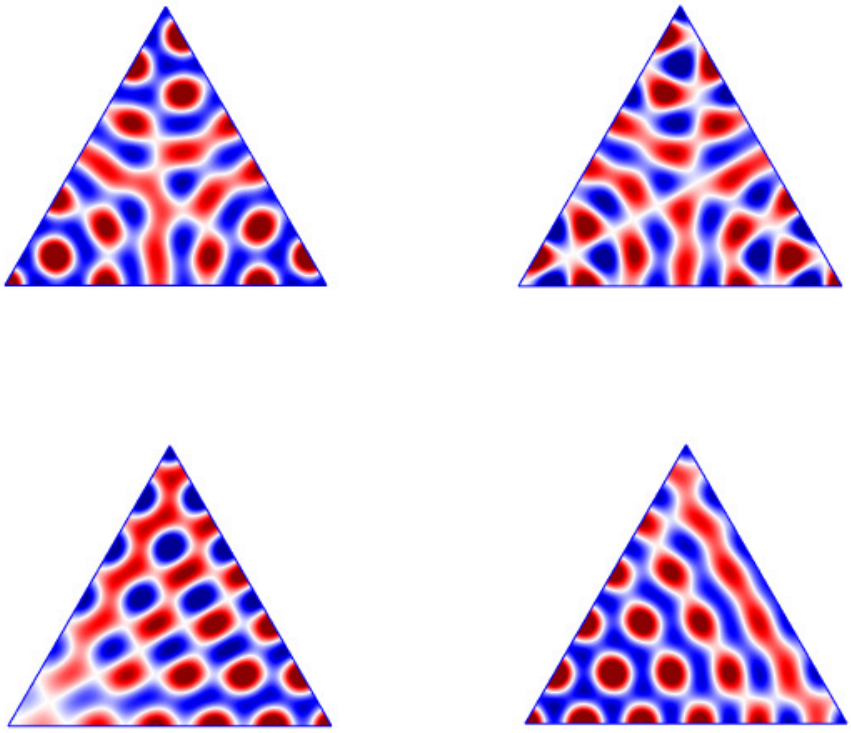}
\caption{Four fold degenerate eigenvalues of equilateral triangle
with eigenfrequencies $\nu_{-11,-19}^2=\nu_{-16,-17}^2=161.778$. }
\label{eigtrian}
\end{figure}

The
eigenfrequencies of equilateral triangle  equal for the Neumann
boundary conditions
\begin{equation}\label{eigstriangle}
\nu_{mn}^2=\omega_{mn}^2/\omega_0^2=\frac{16}{27}(m^2+n^2-mn),
m,n=0, \pm 1, \pm 2, \ldots,
\end{equation}
where $\omega_0$ is defined in Eq. (\ref{eigfreqrect}) with the
following conditions: $m+n$ is a multiple of 3 \cite{Pinsky1980}.
The eigenmodes are of the form
\begin{eqnarray}\label{eigfuntriangle}
   & \psi_{mn}=f_{mn}+f_{m,m-n}+f_{-n,m-n}+f_{-n,-m}+f_{n-m,-m}+f_{n-m,n}&\nonumber\\
&f_{mn}(x,y)=\exp(2\pi i/3)(nx+(2n-m)y/\sqrt{3}).&
\end{eqnarray}
 In Fig. \ref{eigtrian} we
show patterns of the eigenmodes which are four fold degenerate due
to permutation symmetry $m\leftrightarrow n$ and accidental number
degeneracy at $m=-16, n=-17$ and $m=11, n=-19$.

There are also other cases of the four fold degeneracy however
with higher eigenfrequencies with $m=-13, n=-23, m=-23, n=-13$ and
$m=-17, n=-22, m=-22, n=-17$ with the eigenvalue
$\nu_{-13,-23}^2=226.312$.  The group of symmetry $C_{3v}$ of
equilateral triangle consists of rotations $C_3$, mirror
reflections $\sigma_v$  along the diagonals of resonator
\cite{Bir_Pikus}. Tabl. \ref{Tab4} shows the irreducible
representations of each symmetry transformation and their
characters \cite{Bir_Pikus}.
\begin{table}
\caption{The characters of irreducible representations of group
symmetry $C_{3v}$}
\begin{tabular}{|c|c|c|c|c|}
  \hline
$C_{3v}$ & 1 & 2$C_3$ &3$\sigma_v$ & basic modes \\  \hline
$A_1$   &1  & 1     & 1   & $z$ \\
  \hline
$A_2$   & 1 & 1     & -1 &$J_z$ \\
  \hline
$E$ & 2 & -1 & 0 &$x, y$\\
  \hline
 \end{tabular}
 \label{Tab4}
\end{table}
\newpage


%
\end{document}